\def\papertitle{Voice conversion with limited data and limitless data augmentations}
\def\paperauthorA{Olga Slizovskaia}
\def\paperauthorB{Jordi Janer}
\def\paperauthorC{Pritish Chandna}
\def\paperauthorD{Oscar Mayor}

\documentclass[twoside,a4paper]{article}
\usepackage{etoolbox}
\usepackage{dafx_22a}
\usepackage{amsmath,amssymb,amsfonts,amsthm}
\usepackage{euscript}
\usepackage[T1]{fontenc}
\usepackage[utf8]{inputenc}
\usepackage{nimbusserif}
\usepackage{ifpdf}
\usepackage[english]{babel}
\usepackage{caption}
\usepackage{color}

\usepackage{xcolor}
\usepackage{multirow}
\usepackage{siunitx}
\robustify\bfseries
\usepackage{booktabs}
\usepackage{graphics}
\usepackage{balance}
\usepackage{subcaption}

\input glyphtounicode
\ninept

\newcounter{numauth}
\newcounter{listcnt}
\newcommand\authcnt[1]{\ifdefined#1 \stepcounter{numauth} \fi}

\newcommand\addauth[1]{
\ifdefined#1 
\stepcounter{listcnt}
\ifnum \value{listcnt}<\value{numauth}
\appto\authorslist{, #1}
\else
\appto\authorslist{~and~#1}
\fi
\fi}
\authcnt{\paperauthorB}
\authcnt{\paperauthorC}
\authcnt{\paperauthorD}
\authcnt{\paperauthorE}
\authcnt{\paperauthorF}
\authcnt{\paperauthorG}
\authcnt{\paperauthorH}
\authcnt{\paperauthorI}
\authcnt{\paperauthorJ}
\def\authorslist{\paperauthorA}
\addauth{\paperauthorB}
\addauth{\paperauthorC}
\addauth{\paperauthorD}
\addauth{\paperauthorE}
\addauth{\paperauthorF}
\addauth{\paperauthorG}
\addauth{\paperauthorH}
\addauth{\paperauthorI}
\addauth{\paperauthorJ}

\usepackage{times}

\newif\ifpdf
\ifx\pdfoutput\relax
\else
   \ifcase\pdfoutput
      \pdffalse
   \else
      \pdftrue
\fi

\ifpdf 
  \usepackage[pdftex,
    pdftitle={\papertitle},
    pdfauthor={\authorslist},
    pdfsubject={Proceedings of the 25th International Conference on Digital Audio Effects (DAFx20in22)},
    colorlinks=false, 
    bookmarksnumbered, 
    pdfstartview=XYZ 
  ]{hyperref}
  \usepackage[pdftex]{graphicx}
\else 
  \usepackage[dvips]{epsfig,graphicx}
  \usepackage[dvips,
    pdftitle={\papertitle},
    pdfauthor={\authorslist},
    pdfsubject={Proceedings of the 25th International Conference on Digital Audio Effects (DAFx20in22)},
    colorlinks=false, 
    bookmarksnumbered, 
    pdfstartview=XYZ 
  ]{hyperref}
\fi
\usepackage[hypcap=true]{caption}
\title{\papertitle}

\fouraffiliations{
\paperauthorA }
{\href{https://www.voctrolabs.com/}{Voctro Labs S.L.} \\ Barcelona, Spain\\
{\tt \href{olga.slizovskaia@voctrolabs.com}{olga.slizovskaia@voctrolabs.com}}
}
{\paperauthorB }
{\href{https://www.voctrolabs.com/}{Voctro Labs S.L.} \\ Barcelona, Spain\\ {\tt \href{mailto:jordi.janer@voctrolabs.com}{jordi.janer@voctrolabs.com}}
}
{\paperauthorC }
{\href{http://voicemod.net/}{Voicemod} \\Valencia, Spain \\ {\tt \href{mailto:pritish.chandna@voicemod.net}{pritish.chandna@voicemod.net}}
}
{\paperauthorD }
{\href{https://www.voctrolabs.com/}{Voctro Labs S.L.} \\ Barcelona, Spain\\
{\tt \href{oscar.mayor@voctrolabs.com}{oscar.mayor@voctrolabs.com}}
}

\begin{document}
\ifpdf 
  \DeclareGraphicsExtensions{.png,.jpg,.pdf}
\else  
  \DeclareGraphicsExtensions{.eps}
\fi


\maketitle

\begin{abstract}

Applying changes to an input speech signal to change the perceived speaker of speech to a target while maintaining the content of the input is a challenging but interesting task known as Voice conversion (VC). Over the last few years, this task has gained significant interest where most systems use data-driven machine learning models. Doing the conversion in a low-latency real-world scenario is even more challenging constrained by the availability of high-quality data. Data augmentations such as pitch shifting and noise addition are often used to increase the amount of data used for training machine learning based models for this task. In this paper we explore the efficacy of common data augmentation techniques for real-time voice conversion and introduce novel techniques for data augmentation based on audio and voice transformation effects as well. We evaluate the conversions for both male and female target speakers using objective and subjective evaluation methodologies.


\end{abstract}

\section{Introduction}


Voice Conversion (VC) aims at transforming a speech utterance by a source speaker as if it was uttered by a different target speaker. Input speech content shall be  maintained in the converted speech utterance. 
In the recent years we have seen a significant advance and a lots of proposals on how to solve voice conversion problem \cite{sisman2020overview}. Some conventional VC systems employ datasets with parallel utterances by source and target speakers \cite{lorenzo2018voice}, which are difficult and expensive to obtain. Non-parallel VC approaches\cite{zhao2022disentangling, lee2021voicemixer, barbany2020fastvc}, on the other hand, require only datasets of utterances by the target speaker, which is more feasible to obtain, and offer a wider range of practical applications.  VC systems can operate either offline, observing and processing a whole utterance, or in a streaming setting, which can allow for low latency applications in real-time.  
In this paper we focus on any-to-one voice conversion problem under some defined realistic real-time streaming requirements. To meet these requirements our system should process the input audio with a latency under $90$ ms, and it should run at $3$ faster than real-time on a single off-the-shelf CPU. We note that under these constraints, the voice conversion is limited to timbre conversion and pitch-shifting. The prosody of the input signal, which includes the intonation, stress, and rhythm of the speech is very hard to change in low-latency scenario. While there are existing approaches that aim at controlling the prosody for voice conversion \cite{zheng2021prosody}, they do not meet real-time constraints that we have in out work. 

While several systems have been proposed that meet real-time requirements \cite{tacotron2018vc, RTVCCorentinJ, barbany2020fastvc}, and some of them can achieve a convincing and naturally sounding target timbre in the converted speech, most of those systems have additional constraints such and high latency requirements and therefore are not suitable for low-latency streaming use.

Existing approaches for low-latency voice conversion systems include, for example, \cite{saeki2020realtime}: a fully online multi-band system that is able to produce a high-quality output speech at $5$ ms hop size. However, the authors do not discuss if their model achieves a good target speaker similarity and the method can not be evaluated qualitatively or quantitatively as no audio samples are available. Another close approach is AC-VC \cite{ronssin2021acvc} that has only 57.5ms future look ahead and low computational complexity, and, therefore, suitable for real-time and streaming applications. However, this system is not able to alternate the source prosody and achieve convincing target prosody in the converted speech.

Another common constraints for training non-parallel VC systems is a requirement for having a significant amount of target speaker audio, consisting typically of several hours of studio recorded material. Several previous work have addressed this issue by proposing few-shot and zero-shot voice conversion systems, including \cite{arik2018fewsamplesneuralvc, dang2021training, kim2021assemvc}. This systems are usually composed of an encoder and a decoder components which helps to reduce the data constraints for the target timbre modelling. However, for such architectures, if we use too much target speaker data to train the decoder module, and the encoder is not able to fully decompose linguistic content from voice style, the decoder may rely on leakages from the source speech resulting in a perceived source identity in the converted speech. 

In this study, we also use an encoder-decoder voice conversion system and we discuss how the use of audio effects as data augmentation strategies can reduce the amount of target speaker recordings needed, while maintaining the quality of the converted output. We demonstrate that while the amount of available target speaker data does not play a crucial role in achieving high target speaker identity and audio quality, data augmentation techniques focused on timbre transformation and prosody transformation help to reduce source identity leakage and improve target speaker similarity.

\section{Method}

\begin{figure}[t]
  \centering
  \includegraphics[width=\linewidth]{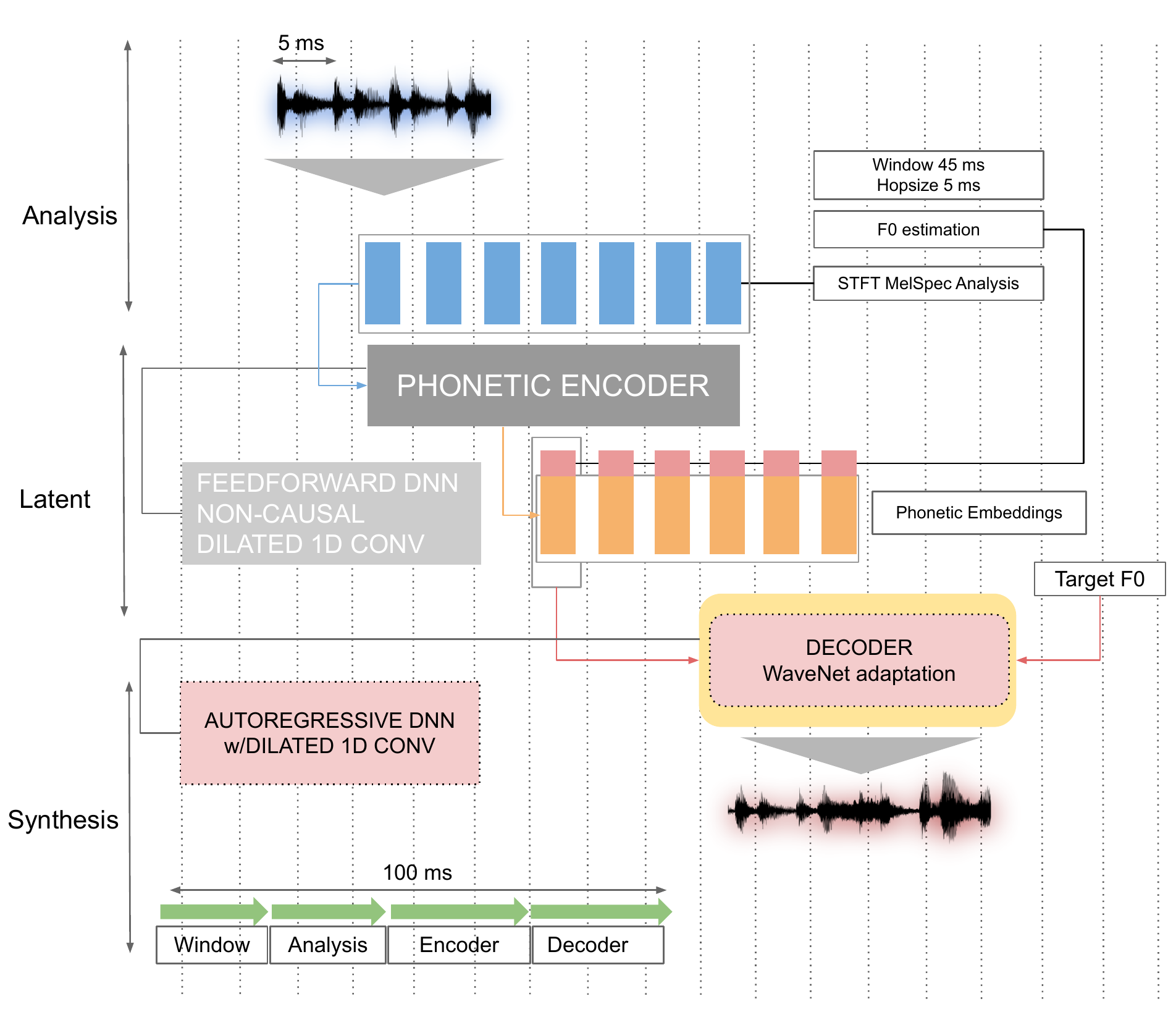}
  \caption{Schematic overview of the baseline voice conversion pipeline.}
  \label{fig:vocov_overview}
\end{figure}

\subsection{Baseline architecture}
The overview of the baseline architecture is shown in Figure~\ref{fig:vocov_overview}. The pipeline consists of two multi-layer
deep neural networks composed of 1D convolutional layers: an encoder and a control-conditioned decoder. The encoder separates or \textit{disentangles} the speaker independent linguistic content from the speaker dependent content like timbre, present in the input signal. The decoder synthesises target speech from the speaker-independent latent representation (phonetic embeddings) and estimated source pitch contour. 

\subsubsection{Encoder}
To help disentangle linguistic content from speaker style, our system uses a pretrained encoder that outputs phonetic embeddings. The encoder is trained as a phoneme classifier on phoneme labels data from over 900 speakers in English. To extract the encoder features, the input audio is converted to a Mel Spectrogram representation (with a window size of $45$ ms and $5$ ms hop-size), and a $120$-dimensional feature vector is taken from the second to the last layer of the trained phoneme classifier. The classifier is composed of nine non-causal dilated 1D convolutional layers and has a total receptive field length of 125 ms which allows for only 65 ms extra latency at the inference time while providing good classification performance.  

For the goal of voice conversion, we expect the encoder features to be speaker-independent and keep only linguistic information from the input speech utterance. Therefore, the same encoder can be used to train multiple decoder models for different target speakers. However, while the phonetic embeddings are an acceptable intermediate representation, it preserves more than just linguistic content from the input speech: they implicitly carry speech rate information and some prosody content hidden in durations of vowels and consonants. The strong dependence on the source prosody is one of the fundamental limitations for real-time voice conversion models. 

\subsubsection{Decoder}

The decoder model is based on the WaveNet architecture \cite{wavenet}, a generative autoregressive model originally proposed for text-to-speech synthesis. In our case, intermediate phonetic representation serves as input to the model, together with pitch controls that are discussed in Section~\ref{controls}.
The decoder predicts a discretised Mixture of Logistic distribution of a discrete transform of a target audio computed at the same rate as phonetic embeddings, with a hop size of $5$ ms. For the decoder architecture we follow the WaveNet vocoder implementation from \cite{blaauw2019data}. 

\subsubsection{Controls}\label{controls}

With the decoder we aim to generate a high quality speech signal that replicates the timbre of the target speaker while maintaining the linguistic content of the input. However, other important components of the speaker identity in speech are pitch and prosody. While there are models that aim for an end-to-end voice conversion and do not separate prosody from linguistic content, they often result in robotically sounding output speech due to network's inability to model the target prosody correctly. To account for this, we use an estimated pitch contour (F0) of the source audio together with a predicted confidence of F0 estimation (that can be seen as a continuous representation for voiced-unvoiced region estimations) as extra controls for the decoder model.  

\subsection{Data}

As the baseline method requires a high level of decomposition of linguistic and acoustic content, we first trained a phonetic encoder on a large scale LibriTTS dataset\cite{zen2019libritts}. To train the encoder, we only used a subset of the main dataset named LibriTTS-360 that contains around 360 hours of audio recordings from 904 speakers. The audio recordings are accompanied with text transcriptions, and the low quality audios and audios with significant background noise are excluded from this subset. All audios have sampling rate of 24 kHz. 

To obtain phonetic transcriptions we used Montreal Forced Aligner \cite{mfa} with CMU US English Dictionary pronunciation dictionary. After the alignment, the recordings containing out-of-dictionary words were filtered out (approx. 5.8\% of the dataset).

The decoder is trained on two proprietary datasets (1 female and 1 male professional voice actors), recorded in a studio with a professional Neumann cardioid microphone with a pop filter and an additional portable voice recording booth. Both actors are recorded in the same conditions, and the distance between speakers and the microphone is approximately 20 cm. The total amount of recordings is 3.5 hours for the female dataset and 1.8 hours for the male dataset. All audios are converted to 16bit single-channel at 24kHz sampling rate for this study. For each target, three clean baseline datasets are created for training according to the total audio duration in the dataset: 15 minutes, 1 hour and Full datasets.

\subsection{Data augmentations}

We study several types of data augmentation (DA) techniques in this paper. 

The first group includes conventional data augmentations techniques based on adding environmental and white Gaussian noises to the source data or source controls:
\begin{itemize}
    \item Noise augmentations that consists in creating mixtures of clean target audio with environmental noise samples from the WHAM!\cite{wham} dataset (denoted as Noisy in the experiments);
    \item Additive white Gaussian random noise applied to the pitch controls (denoted as NoisyF0);
    \item Pitch controls smoothing with an additive smooth white Gaussian noise (denoted as NoisyF0-SM).
\end{itemize}

The second group is focused on applying audio effects to the source data. We used speech and vocal transformation effects for this step, processing the original audio with the SoX \footnote{\url{http://sox.sourceforge.net/}} library and the VoTrans library \cite{bonada2008wide, mayor2011audio}.  VoTrans implements a pulse-based analysis and synthesis method that allows for both pitch-shifting with formant preservation (keeping identity), and gender or voice character change, by manipulating the spectral envelope of individual voice pulses in the frequency domain. Augmentations here consist of:

\begin{itemize}
    \item Pitch-shift data augmentations with SoX (denoted as SoX in the experiments);
    \item Spectral envelope transformation with pitch-shift, that controls a random variable for the type and amount of timbre modification via spectral envelope warping (denoted as VoTrans);
    \item A combination of the above two together with additive noise over the pitch controls (denoted as NoisyF0-VT-SoX).
\end{itemize}

In addition, the encoder is trained using spectral masking data augmentations but its effect is not discussed in this study.

\subsection{Implementation and training details}

For offline DAs (Noisy, SoX, VoTrans and NoisyF0-VT-SoX which would be slow to use as on-the-fly augmentations) we first preprocess the target datasets and store the transformed audio files for the further use in training. Instead, the online DAs (NoisyF0 and NoisyF0-SM) are implemented as a parameterised TensorFlow function. Only training data is augmented while validation data is kept clean.

The augmentation details are as follows:

\begin{itemize}
    \item Noisy: for each source clean audio we select 5 random noise segments and make a noisy mixture at a randomly selected source-to-noise ratio drawn from a uniform distribution $\mathrm{SNR} \sim \mathcal{U}(4,\, 12) $ dB;
    \item NoisyF0: at each training step a random noise is added to each value of the estimated source fundamental frequency. The noise values are drawn from a Gaussian distribution $x_{f0} \sim \mathcal{N}(0,\, 0.25)$. The F0 confidence values are augmented by random values drawn from a steeper distribution with the same mean $x_{conf} \sim \mathcal{N}(0,\, 0.215)$;
    \item NoisyF0-SM: at each training step we first pick a random smoothing window size from a uniform distribution $\mathrm{S} \sim \mathcal{U}(100,\, 300) $ ms and compute a smooth F0 curve. Then, white Gaussian noise is sampled as for NoiseF0, smoothed again with a smaller window size and added to get the final augmented F0 curve;
    \item SoX: for each clean source audio we generate 10 augmented audios each of those is pitch-shifted by a random number of semitones drawn from a Gaussian distribution $ p \sim min (\mathcal{N}(0,\, 3), 8)$ limited by 8 semitones at most. We observe that doing pitch shift with the SoX library with values above 8 semitones significantly degrade output quality and speech intelligibility. We use the pySoX \cite{bittner2016pysox} library for the data preprocessing;
    \item VoTrans: for each clean source audio we generate 10 augmented audios. First a pitch shift value is sampled from a uniform distribution  $p \sim \mathcal{U}(-12,\, 12) $ semitones. Then, we expand or compress the formants of the source spectrum based on the amount of pitch shift applied.
    \item NoisyF0-VT, NoisyF0-VT-SoX: this DAs represent combinations of NoisyF0, VoTrans and SoX augmentations.
\end{itemize}

First, we train baseline models on the clean datasets for each target. Then, we train a corresponding model for each type of data augmentations considered in this study. All models are trained for 500k iterations (approximately 10 hours on a single GPU).

\section{Evaluation and results}

Evaluation of voice conversion models is a highly subjective and complex task \cite{sisman2020overview, vcchallenge2016} since we need not only to evaluate the overall quality of the synthesised audio, but also how well the linguistic content is preserved and the ability of the model to capture the main aspects of the target speaker identity and change the identity in the source sample. For the second part, one possible approach is based on computing a distance between speaker embeddings obtained from a large-scale speaker verification model, which is commonly used for training \cite{zhao2022disentangling, lee2021voicemixer, barbany2020fastvc, arik2018fewsamplesneuralvc, dang2021training} but might not be reliable for evaluation yet. We use a two-fold evaluation approach that combines objective evaluations of mean opinion score (MOS) obtained with a NISQAv2 model \cite{nisqa} and subjective evaluations from user listening tests.

For objective evaluation we use a manually curated set of 21 audio samples taken from different datasets: 8 samples from the CommonVoice dataset\cite{commonvoice2020}, 2 samples from the English Dialect dataset\cite{englishdialiects2020}, and 11 samples recorded or taken from available datasets selected in order to cover diverse input conditions, both in terms of linguistic content such as different languages, and acoustic recording conditions. Overall results are shown in Table~\ref{tab:results}, and we discuss them further in the following subsections. Samples from the following experiments trained with different DAs on Full datasets are available online at {\url{https://voctrolabs.notion.site/Audio-examples-for-data-augmentation-experiments-93caf428c1a646889cff51ad6f24181d}.

\begin{table*}[ht!]
  \caption{Results of objective and subjective evaluations for different targets, dataset sizes and data augmentation scenarios.}
  \label{tab:results}
  \centering
  \sisetup{table-format=2.2,round-mode=places,round-precision=2,table-number-alignment=center,detect-weight=true}
  \begin{tabular}{llSSSSSSSSS[table-format=1.2]}\toprule
            score & dataset &\multicolumn{3}{c}{\textbf{15 minutes}}&\multicolumn{3}{c}{\textbf{1 hour}}&\multicolumn{3}{c}{\textbf{Full data}}
            \\ \cmidrule(lr){3-5}\cmidrule(lr){6-8}\cmidrule(r){9-11}
              & method / gender       & all & m & f & all & m & f & all & m & f \\\midrule
            NISQA    & Clean & 3.77 & 4.19 & 3.34 & 3.70 & 3.99 & 3.42 & 3.70 & 4.00 & 3.39\\
                     & Noisy & 2.89 & 2.95 & 2.84 & 3.09 & 3.12 & 3.06 & 2.80 & 3.29 & 2.31\\
                     & SoX & 3.38 & 3.72 & 3.05 & 3.38 & 3.31 & 3.44 & 3.39 & 3.17 & 3.60\\
                     & VoTrans & 3.64 & 3.83 & 3.44 & 3.87 & 3.97 & 3.77 & 3.70 & 3.83 & 3.56\\
                     & NoisyF0 & 3.84 & 4.28 & 3.40 & 3.71 & 4.16 & 3.26 & 3.75 & 3.90 & 3.59\\
                     & NoisyF0-SM & 3.62 & 3.84 & 3.40 & 3.53 & 3.87 & 3.18 & 3.57 & 3.79 & 3.35\\
                     & NoisyF0-VT & 3.78 & 3.91 & 3.64 & 3.34 & 3.38 & 3.31 & 3.40 & 3.57 & 3.23\\
                     & NoisyF0-VT-SoX & 3.26 & 3.62 & 2.89 & 2.95 & 3.17 & 2.73 & 3.25 & 3.33 & 3.17\\\midrule
            
            MOS     & Clean & 2.45 & 3.18 & 2.02 & 2.65 & 2.86 & 2.42 & 2.66 & 3.00 & 2.21\\
                    & SoX & 1.98 & 2.65 & 1.58 & 2.18 & 2.47 & 1.87 & 2.16 & 2.41 & 1.82\\
                    & VoTrans & 2.05 & 2.33 & 1.88 & 2.61 & 2.92 & 2.28 & 2.49 & 2.80 & 2.08\\
                    & NoisyF0 & 1.59 & 1.86 & 1.43 & 2.13 & 2.77 & 1.43 & 2.27 & 2.57 & 1.86\\
                    & NoisyF0-VT-SoX & 1.34 & 2.17 & 0.85 & 1.64 & 2.36 & 0.85 & 1.83 & 2.25 & 1.27\\ \midrule
                    
            SIM     & Clean & n/a & n/a & n/a & 1.85 & 2.06 & 1.70 & 2.03 & 2.28 & 1.81\\
                    & VoTrans & n/a & n/a & n/a & 1.90 & 2.11 & 1.74 & 1.96 & 2.01 & 1.93\\
                    & NoisyF0 & n/a & n/a & n/a & 1.81 & 2.27 & 1.46 & 1.95 & 2.26 & 1.68\\
                    & NoisyF0-VT & n/a & n/a & n/a & 1.83 & 2.16 & 1.57 & 2.03 & 2.33 & 1.78
            \\\bottomrule
    \end{tabular}
  
\end{table*}

\subsection{NISQA MOS score analysis}

While several objective metrics exists for speech quality and intelligibility assessment (such as PESQ \cite{pesq}, STOI \cite{stoi}), they usually require a clean reference signal. While this approach works well for, for example, denoising, the need of a reference audio is a severe limitation in voice conversion task because that signal is usually unavailable. Several non-intrusive alternatives have been proposed recently, including Non-intrusive Objective Speech Quality Assessment (NISQA) \cite{nisqa} that is used in this work.

In Table~\ref{tab:results}, it can be observed that training on clean data yields higher scores for the male target when all available data is used for training ($4.0$ MOS), while NoisyF0 augmented model outperforms non-augmented model on 15m and 1h datasets ($4.28$ and $4.16$ MOS). Clean and NoisyF0 is followed by VoTrans and NoisyF0-VT data augmentations which yield slightly lower scores. Interestingly, MOS scores decrease for male target as more data is available for both clean datasets and augmented datasets in all cases except Noisy data augmentations as shown in Fig~\ref{fig:nisqa_male}. 

For the female target, the overall quality is lower compared to the male target. We do not observe a consistent quality improvement related to the dataset size. For the female case, NoisyF0 models do not yield a clear improvement over the clean data baseline even though this model outperforms the baseline for 15m and Full datasets. The performance of the NoisyF0-SM and NoisyF0-VT models is comparable to that of the baseline. Nevertheless, the highest scores are obtained using VoTrans augmentations ($3.77$ MOS on 1h dataset vs. $3.42$ MOS for the corresponding baseline), and VoTrans augmentations outperform the baseline consistently for all dataset sizes. SoX augmentations result in better performance for larger datasets (1h, Full), while NoisyF0-VT-SoX is consistently inferior to the clean data baselines.

For all targets and datasets, augmentations with noise samples result in the overall lowest NISQA scores and oversmoothed re-synthesised audio. The damaging effect seems to be similar in the female case (1.08 MOS loss on the Full dataset) and in the male case ($1.24$ MOS loss on the 15m dataset). Furthermore, SoX, NoisyF0-SM and NoisyF0-VT-SoX do not yield higher scores compared to the clean data (see Appendix~\ref{appendix} for MOS score visualizations). 

\subsection{Subjective listening tests}

For the listening tests, we select a subset of audio samples that have a better input quality and setup an online MUSHRA listening tests using the beaqleJS framework \cite{beaqlejs}. We obtain a total of 1482 scores from 12 participants however, we only get 128 scores for the samples from models trained on the smallest datasets (15m), and those scores are excluded from the analysis. Thus, we only analyse results for the models trained on two subsets for each target: 1h and Full dataset. 

The participants are asked two evaluate two aspects of test samples compared to the reference audio: overall audio quality (includes intelligibility and naturalness), and target speaker similarity. Curiously, the average score obtained from the listening test experiments is $1.5$ points lower compared to the NISQA model: $3.6$ NISQA mean score vs. $2.1$ mean score from the listening test results. We now analyse the listening test results in more details.

\subsubsection{MOS scores}

For all targets, the highest scores are obtained from the models trained on clean data, except for the male target model trained on the 1h dataset with VoTrans augmentations. Similar to the objective evaluation, adding more data do not improve the performance for neither the male or the female voice. We observe a descend in quality for all considered augmentation techniques, that is also shown in Figure~\ref{fig:mos_all}, Figure~\ref{fig:mos_female} and Figure~\ref{fig:mos_male}. Most noticeably, the MOS scores decrease for the models trained on NoisyF0-VT-SoX augmented data, especially in the female case, that only reaches $0.85$ MOS on 15m and 1h datasets. The results for the models trained with only NoisyF0 augmentations also worsen the performance, although not as dramatically as in the combined augmentations case. It is followed by SoX augmented models that make less damage than the previous cases, and, finally VoTrans augmented models have the best overall quality among the augmented models. 

\subsubsection{Target similarity scores}

In this section we discuss target similarity scores (SIM) obtained from the listening tests. Please, note that we will only discuss the results obtained from 1h and Full datasets as there are not enough ratings for reliable evaluation of 15m dataset models.

Similarly to the objective and subjective MOS scores, there is no clear foundation that adding more data to the target voice dataset significantly improve target speaker similarity. However, if we only consider models trained on 1h and Full datasets, the similarity scores raise accordingly from $1.85$ SIM to $2.03$ SIM for the clean data baselines. 

We observe that the SIM scores improve for the male target then trained on NoisyF0-VT augmented dataset for all dataset sizes. However, this trend does not hold for the female voice. NoisyF0 augmentations perform comparably to the baseline, resulting in $0.21$ points improvement for the male target trained on the 1h dataset, while decreasing the female similarity score by $0.24$ points for the same dataset size. VoTrans augmentations also perform comparably to the baseline, improving the average similarity scores by $0.05$ on the 1h dataset while decreasing them by $0.07$ for the Full dataset. Overall, we observe that the male voice target benefits from all augmentations on the 1h dataset in terms of the target similarity. 





\section{Conclusions}

In this paper, we study and measure the impact that target voice datasets have on the performances of a voice conversion model. We explore two aspects related to the target voice dataset: its size and a possible employment of different types of data augmentations.   

First, through the extensive experimentation, we show that simply increasing the amount of target data used for training does not provide a significant performance boost in neither output audio quality nor target speaker identity.  
Second, we compare effectiveness of a set of audio data augmentation techniques applied to the voice conversion problem. On one hand, we show that the excess of data augmentations, especially those that lead to significant input quality degradation (such as mixtures with ambient noises, extreme SoX augmentations and NoisyF0-VT-SoX), have a negative impact on converted audio quality. On the other hand, the use of more carefully designed augmentation techniques (such as timbre transformation with VoTrans and modifying prosody with NoisyF0) may be helpful in achieving more realistic target speaker identity and better audio quality. We also note that conversion to a male speaker was more effective than conversion to a female speaker even though the quality and quantity of training data for both datasets was similar.

\section{Acknowledgements}

The authors would like to thank Jordi Bonada and Merlijn Blaauw for kindly providing the non open-sourced baseline system architecture and discussions. We also thank Julien Arres and Federico López for their helpful feedback. This work was partially supported by the Spanish Ministry of Science and Innovation under the Torres Quevedo program (PTQ2020-011269).  

\nocite{*}
\bibliographystyle{IEEEbib}
\bibliography{DAFx22_tmpl} 

\begin{thebibliography}{10}

\bibitem{sisman2020overview}
Berrak Sisman, Junichi Yamagishi, Simon King, and Haizhou Li,
\newblock ``An overview of voice conversion and its challenges: From
  statistical modeling to deep learning,''
\newblock {\em IEEE Trans. Audio, Speech, Lang. Process.}, vol. 29, pp.
  132--157, 2020.

\bibitem{lorenzo2018voice}
Jaime Lorenzo-Trueba, Junichi Yamagishi, Tomoki Toda, Daisuke Saito, Fernando
  Villavicencio, Tomi Kinnunen, and Zhenhua Ling,
\newblock ``The voice conversion challenge 2018: Promoting development of
  parallel and nonparallel methods,''
\newblock {\em arXiv preprint arXiv:1804.04262}, 2018.

\bibitem{zhao2022disentangling}
Xintao Zhao, Feng Liu, Changhe Song, Zhiyong Wu, Shiyin Kang, Deyi Tuo, and
  Helen Meng,
\newblock ``{Disentangling Content and Fine-grained Prosody Information via
  Hybrid ASR Bottleneck Features for Voice Conversion},''
\newblock in {\em Proc. ICASSP}, 2022.

\bibitem{lee2021voicemixer}
Sang-Hoon Lee, Ji-Hoon Kim, Hyunseung Chung, and Seong-Whan Lee,
\newblock ``{VoiceMixer: Adversarial Voice Style Mixup},''
\newblock in {\em Proc. NeurIPS}, 2021.

\bibitem{barbany2020fastvc}
Oriol Barbany and Milos Cernak,
\newblock ``{FastVC: Fast Voice Conversion with non-parallel data},''
\newblock in {\em Joint Workshop for the Blizzard Challenge and Voice
  Conversion Challenge 2020}, 2020.

\bibitem{zheng2021prosody}
Zheng Lian, Rongxiu Zhong, Zhengqi Wen, Bin Liu, and Jianhua Tao,
\newblock ``Towards fine-grained prosody control for voice conversion,''
\newblock in {\em 12th International Symposium on Chinese Spoken Language
  Processing (ISCSLP)}, 2021.

\bibitem{tacotron2018vc}
Ye~Jia, Yu~Zhang, Ron~J. Weiss, Quan Wang, Jonathan Shen, Fei Ren, Zhifeng
  Chen, Patrick Nguyen, Ruoming Pang, Ignacio~Lopez Moreno, and Yonghui Wu,
\newblock ``Transfer learning from speaker verification to multispeaker
  text-to-speech synthesis,''
\newblock in {\em Proc. NeurIPS}, 2018, p. 4485–4495.

\bibitem{RTVCCorentinJ}
Corentin Jemine,
\newblock ``Real-time voice cloning,''
  \url{https://github.com/CorentinJ/Real-Time-Voice-Cloning}, 2019.

\bibitem{saeki2020realtime}
Takaaki Saeki, Yuki Saito, Shinnosuke Takamichi, and Hiroshi Saruwatari,
\newblock ``Real-time, full-band, online dnn-based voice conversion system
  using a single cpu.,''
\newblock in {\em Proc. Interspeech}, 2020, pp. 1021--1022.

\bibitem{ronssin2021acvc}
Damien Ronssin and Milos Cernak,
\newblock ``{AC-VC: Non-parallel Low Latency Phonetic Posteriorgrams Based
  Voice Conversion},''
\newblock {\em arXiv preprint arXiv:2111.06601}, 2021.

\bibitem{arik2018fewsamplesneuralvc}
Sercan Arik, Jitong Chen, Kainan Peng, Wei Ping, and Yanqi Zhou,
\newblock ``Neural voice cloning with a few samples,''
\newblock {\em Proc. NeurIPS}, vol. 31, 2018.

\bibitem{dang2021training}
Trung Dang, Dung Tran, Peter Chin, and Kazuhito Koishida,
\newblock ``{Training Robust Zero-Shot Voice Conversion Models with
  Self-supervised Features},''
\newblock in {\em Proc. ICASSP}, 2022.

\bibitem{kim2021assemvc}
Kang-Wook Kim, Seung-Won Park, Junhyeok Lee, and Myun-Chul Joe,
\newblock ``{Assem-VC: Realistic Voice Conversion by Assembling Modern Speech
  Synthesis Techniques},''
\newblock in {\em Proc. ICASSP}, May 2022.

\bibitem{wavenet}
Aäron van~den Oord, Sander Dieleman, Heiga Zen, Karen Simonyan, Oriol Vinyals,
  Alexander Graves, Nal Kalchbrenner, Andrew Senior, and Koray Kavukcuoglu,
\newblock ``{WaveNet: A Generative Model for Raw Audio},''
\newblock in {\em Arxiv}, 2016.

\bibitem{blaauw2019data}
Merlijn Blaauw, Jordi Bonada, and Ryunosuke Daido,
\newblock ``Data efficient voice cloning for neural singing synthesis,''
\newblock in {\em Proc. ICASSP}.

\bibitem{zen2019libritts}
H.~Zen, V.~Dang, R.~Clark, Y.~Zhang, R.~J. Weiss, Y.~Jia, Z.~Chen, and Y.~Wu,
\newblock ``{LibriTTS: A Corpus Derived from LibriSpeech for Text-to-Speech},''
\newblock in {\em Proc. Interspeech}, Sept. 2019.

\bibitem{mfa}
Michael McAuliffe, Michaela Socolof, Sarah Mihuc, Michael Wagner, and Morgan
  Sonderegger,
\newblock ``{Montreal Forced Aligner: Trainable Text-Speech Alignment Using
  Kaldi},''
\newblock in {\em Proc. Interspeech}, 2017, vol. 2017, pp. 498--502.

\bibitem{wham}
Gordon Wichern, Joe Antognini, Michael Flynn, Licheng~Richard Zhu, Emmett
  McQuinn, Dwight Crow, Ethan Manilow, and Jonathan~Le Roux,
\newblock ``{WHAM!: Extending Speech Separation to Noisy Environments},''
\newblock in {\em Proc. Interspeech}, 2019, pp. 1368--1372.

\bibitem{bonada2008wide}
Jordi Bonada,
\newblock ``Wide-band harmonic sinusoidal modeling,''
\newblock in {\em 11th International Conference on Digital Audio Effects DAFx},
  2008, vol.~8, pp. 265--272.

\bibitem{mayor2011audio}
Oscar Mayor, Jordi Bonada, and Jordi Janer,
\newblock ``Audio transformation technologies applied to video games,''
\newblock in {\em Audio Engineering Society Conference: 41st International
  Conference: Audio for Games}, 2011.

\bibitem{bittner2016pysox}
Rachel Bittner, Eric Humphrey, and Juan Bello,
\newblock ``Pysox: Leveraging the audio signal processing power of sox in
  python,''
\newblock in {\em Proc. ISMIR}, 2016,
\newblock {Late Breaking and Demo Papers}.

\bibitem{vcchallenge2016}
Mirjam Wester, Zhizheng Wu, and Junichi Yamagishi,
\newblock ``{Analysis of the Voice Conversion Challenge 2016 Evaluation
  Results},''
\newblock in {\em Proc. Interspeech}, 2016, vol. 2016.

\bibitem{nisqa}
Gabriel Mittag, Babak Naderi, Assmaa Chehadi, and Sebastian M{\"o}ller,
\newblock ``{NISQA: A Deep CNN-Self-Attention Model for Multidimensional Speech
  Quality Prediction with Crowdsourced Datasets},''
\newblock in {\em Proc. Interspeech}, 2021, pp. 2127--2131.

\bibitem{commonvoice2020}
R.~Ardila, M.~Branson, K.~Davis, M.~Henretty, M.~Kohler, J.~Meyer, R.~Morais,
  L.~Saunders, F.~M. Tyers, and G.~Weber,
\newblock ``Common voice: A massively-multilingual speech corpus,''
\newblock in {\em Proceedings of the 12th Conference on Language Resources and
  Evaluation (LREC 2020)}, 2020, pp. 4211--4215.

\bibitem{englishdialiects2020}
Isin Demirsahin, Oddur Kjartansson, Alexander Gutkin, and Clara Rivera,
\newblock ``{Open-source Multi-speaker Corpora of the English Accents in the
  British Isles},''
\newblock in {\em Proceedings of The 12th Language Resources and Evaluation
  Conference (LREC)}, Marseille, France, May 2020, pp. 6532--6541, European
  Language Resources Association (ELRA).

\bibitem{pesq}
Antony~W Rix, John~G Beerends, Michael~P Hollier, and Andries~P Hekstra,
\newblock ``{Perceptual evaluation of speech quality (PESQ) - a new method for
  speech quality assessment of telephone networks and codecs},''
\newblock in {\em Proc. ICASSP}, 2001, vol.~2, pp. 749--752.

\bibitem{stoi}
Cees~H Taal, Richard~C Hendriks, Richard Heusdens, and Jesper Jensen,
\newblock ``An algorithm for intelligibility prediction of time-frequency
  weighted noisy speech,''
\newblock {\em IEEE Trans. Audio, Speech, Lang. Process.}, vol. 19, no. 7, pp.
  2125--2136, 2011.

\bibitem{beaqlejs}
Sebastian Kraft and Udo Z{\"o}lzer,
\newblock ``{BeaqleJS}: {HTML5} and {JavaScript} based framework for the
  subjective evaluation of audio quality,''
\newblock in {\em Linux Audio Conference, Karlsruhe, DE}, 2014.

\bibitem{bonada2014audio}
Jordi Bonada~Sanjaume,
\newblock ``Audio signal transforming by utilizing a computational cost
  function,'' Apr. 2014,
\newblock US Patent 8,706,496.

\end{thebibliography}

\section{Appendix: Results visualization}
\label{appendix}

In this Appendix we provide boxplot visualization of objective and subjective evaluation results for proposed augmentation schemes on different targets and different dataset sizes.

\begin{figure*}
  \centering
    \begin{subfigure}{0.65\columnwidth}
         \includegraphics[width=\columnwidth]{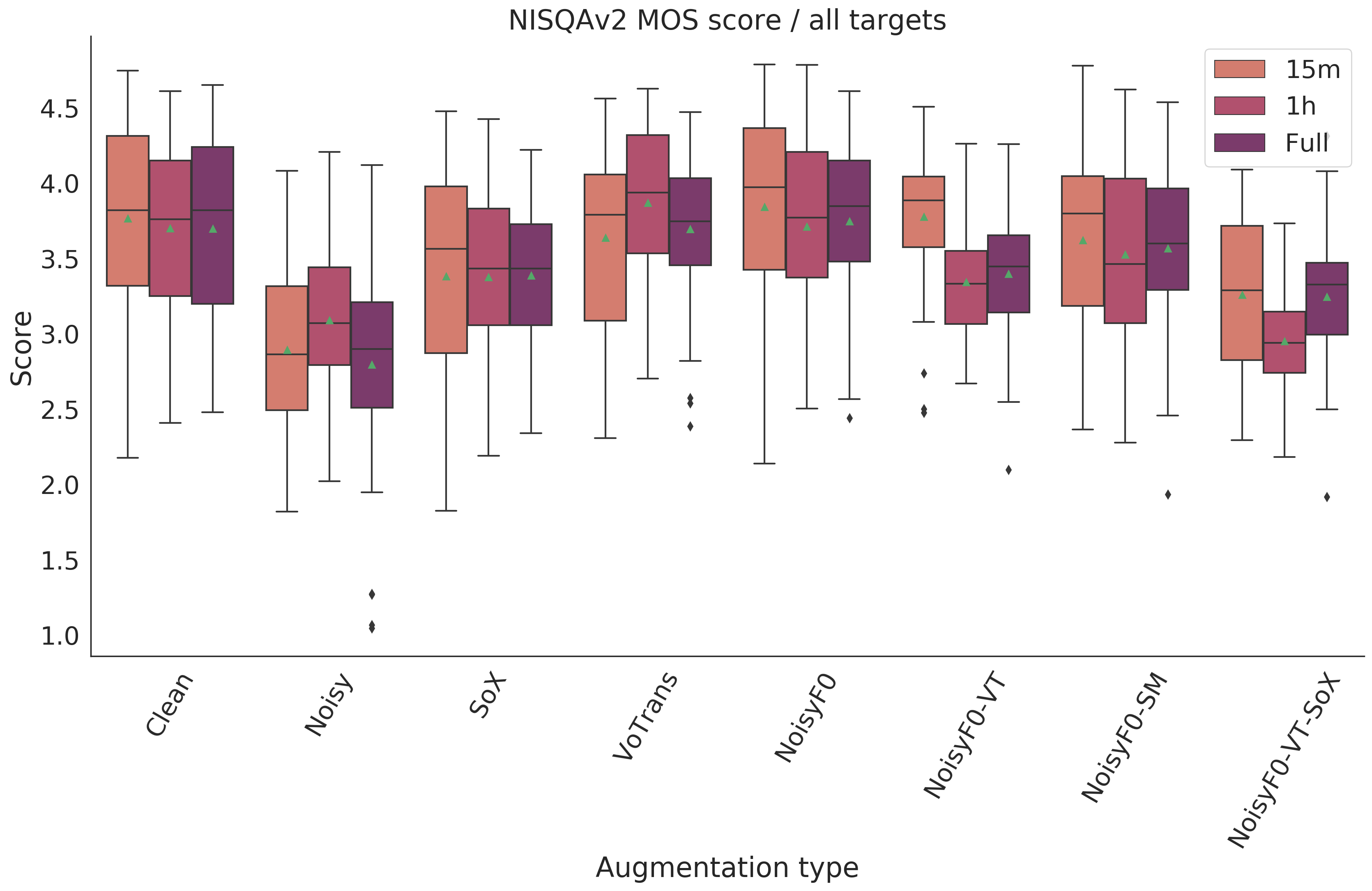}
         \caption{NISQAv2 results for all targets.}
         \label{fig:nisqa_all}
     \end{subfigure} \hfill
    \begin{subfigure}{0.65\columnwidth}
         \includegraphics[width=\columnwidth]{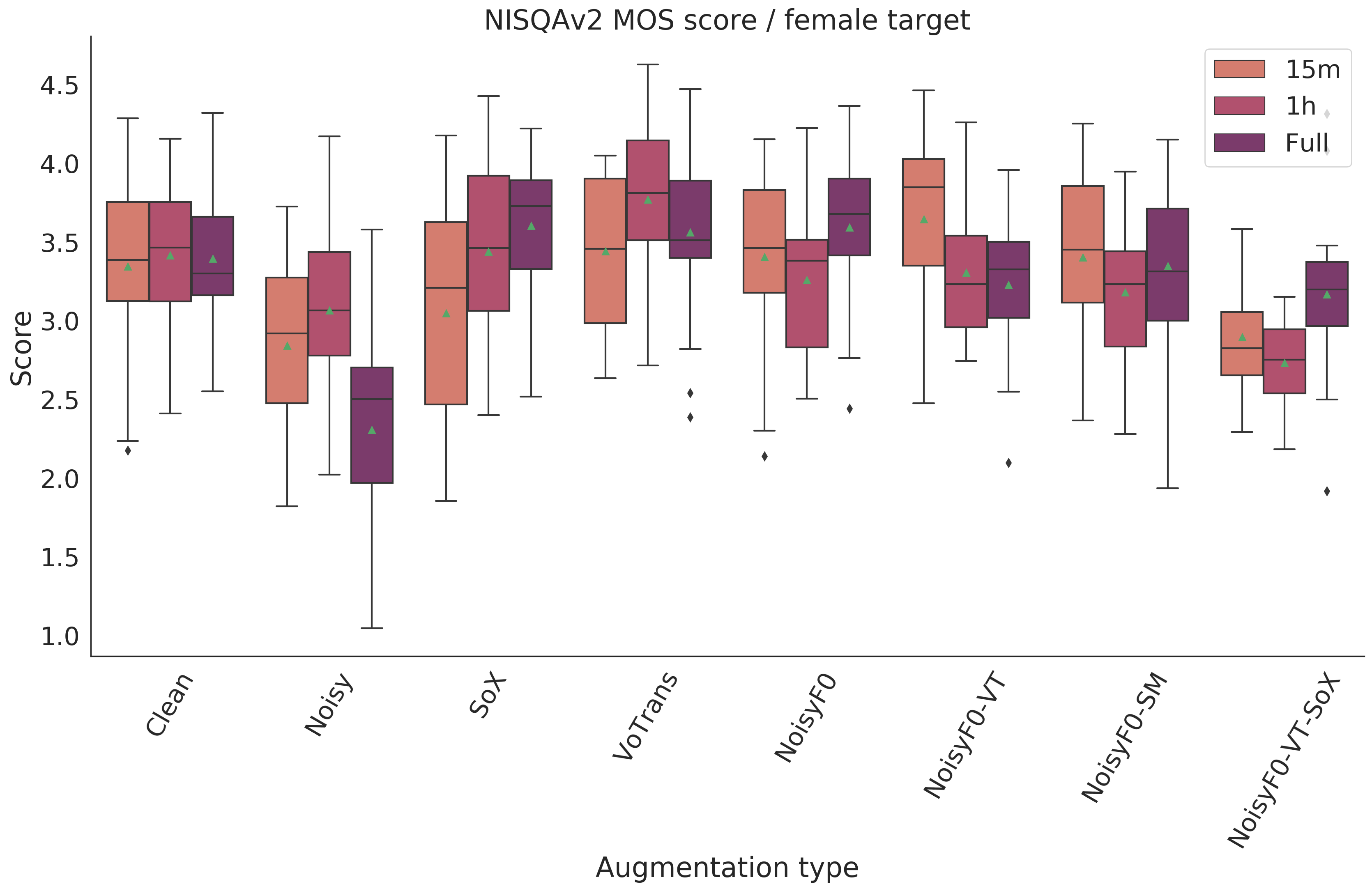}
         \caption{NISQAv2 results for the female target.}
         \label{fig:nisqa_female}
     \end{subfigure} \hfill
    \begin{subfigure}{0.65\columnwidth}
         \includegraphics[width=\columnwidth]{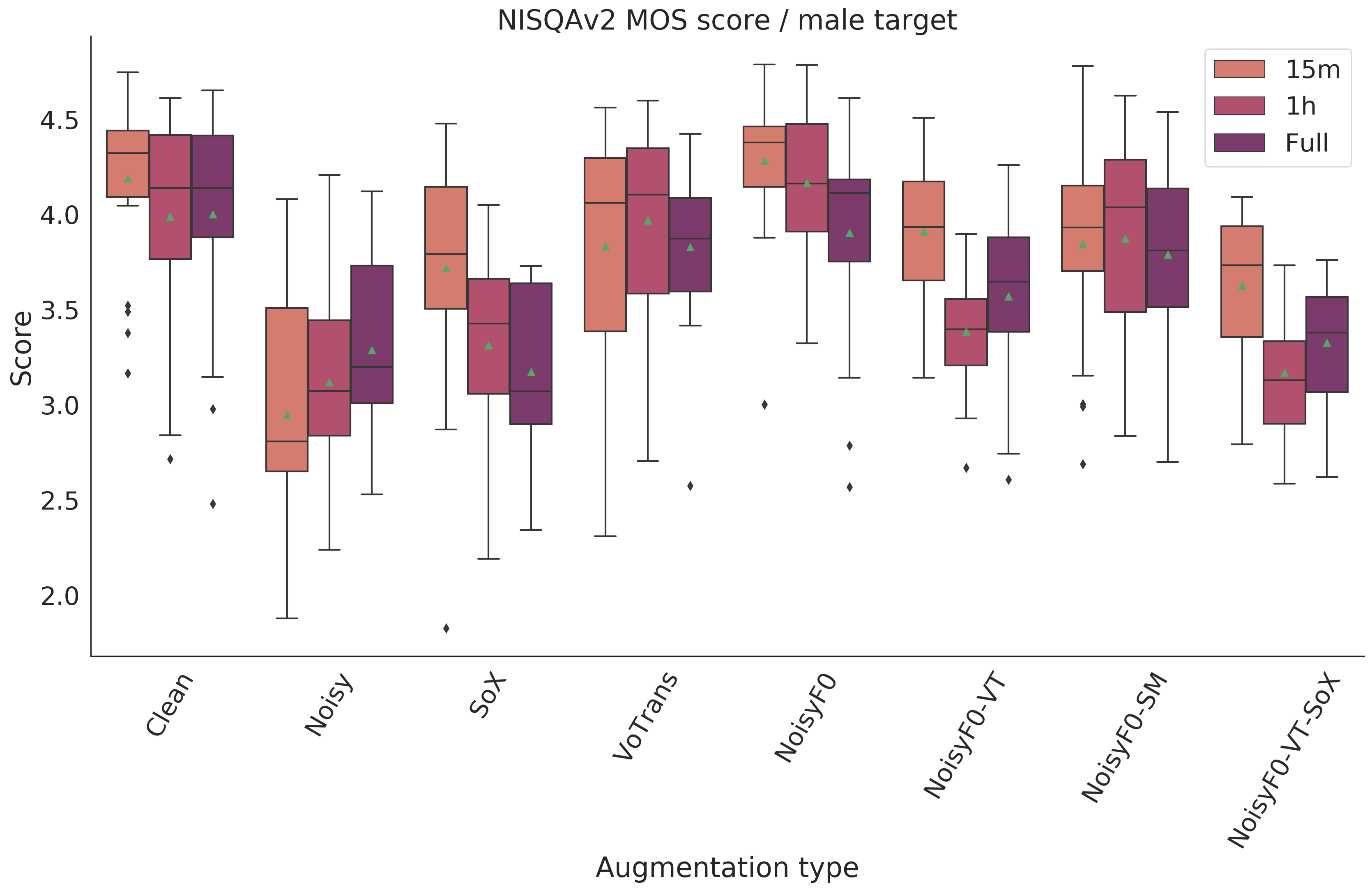}
         \caption{NISQAv2 results for the male target.}
         \label{fig:nisqa_male}
     \end{subfigure} \hfill
  \caption{NISQAv2 objective MOS results for all data augmentations schemes and all targets. }
  \label{nisqa}
\end{figure*}

\begin{figure*}
  \centering
    \begin{subfigure}{0.65\columnwidth}
         \includegraphics[width=\columnwidth]{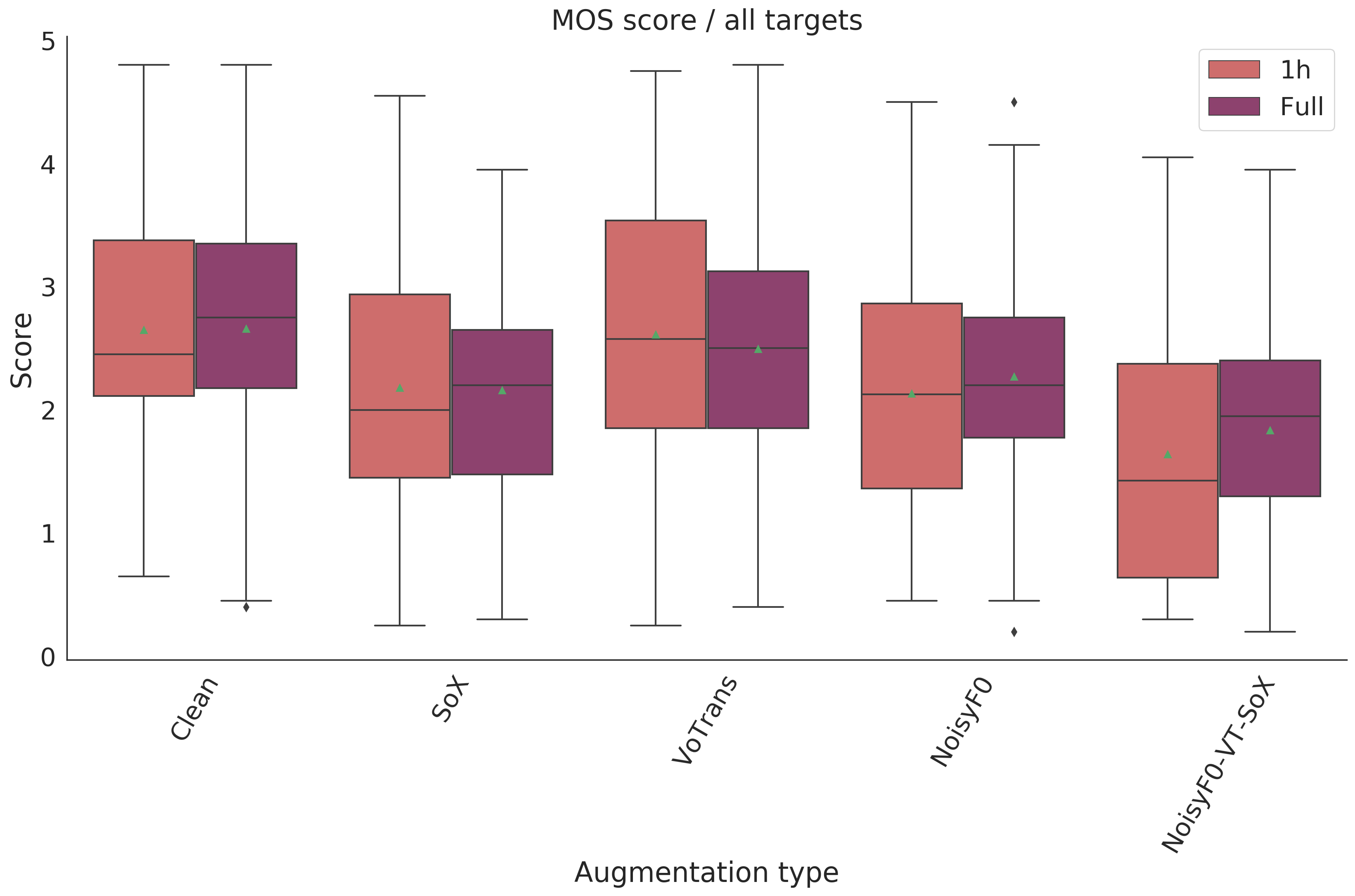}
         \caption{MOS subjective listening test results for all targets.}
         \label{fig:mos_all}
     \end{subfigure} \hfill
    \begin{subfigure}{0.65\columnwidth}
         \includegraphics[width=\columnwidth]{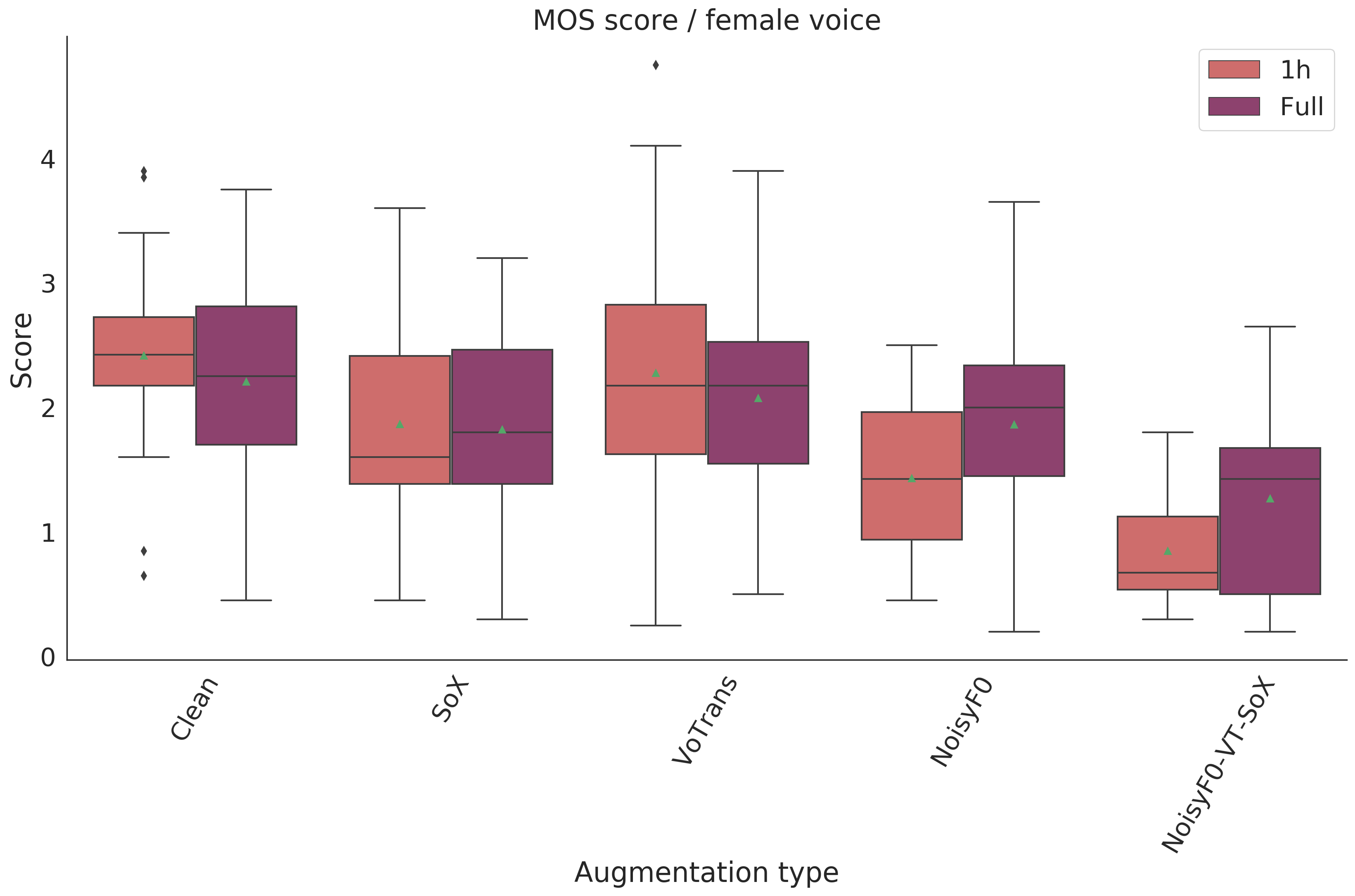}
         \caption{MOS subjective listening test results for the female target.}
         \label{fig:mos_female}
     \end{subfigure} \hfill
    \begin{subfigure}{0.65\columnwidth}
         \includegraphics[width=\columnwidth]{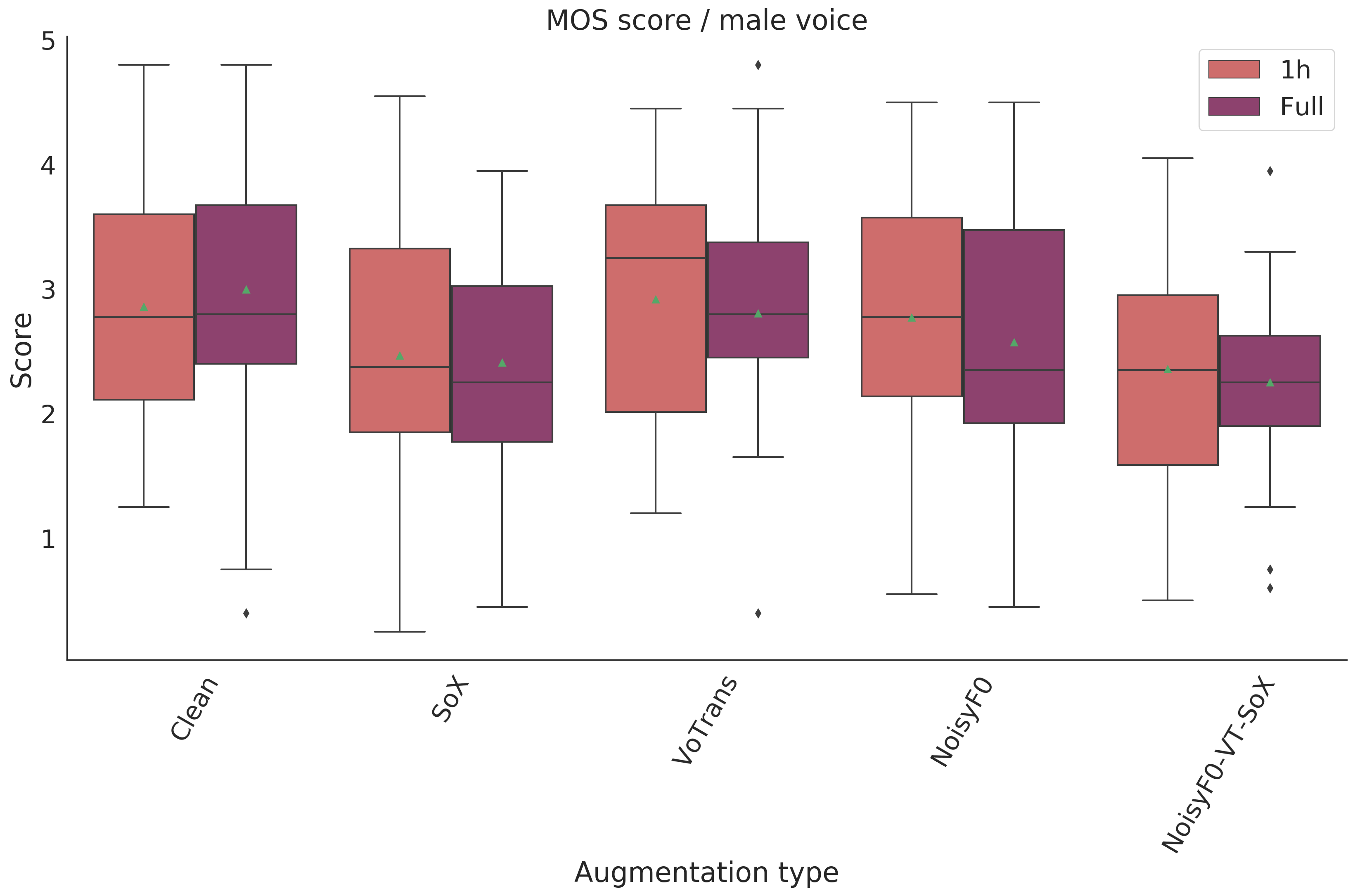}
         \caption{MOS subjective listening test results for the male target.}
         \label{fig:mos_male}
     \end{subfigure} \hfill
  \caption{MOS subjective listening test results for all data augmentations schemes and all targets. }
  \label{mos}
\end{figure*}

\begin{figure*}
  \centering
    \begin{subfigure}{0.65\columnwidth}
         \includegraphics[width=\columnwidth]{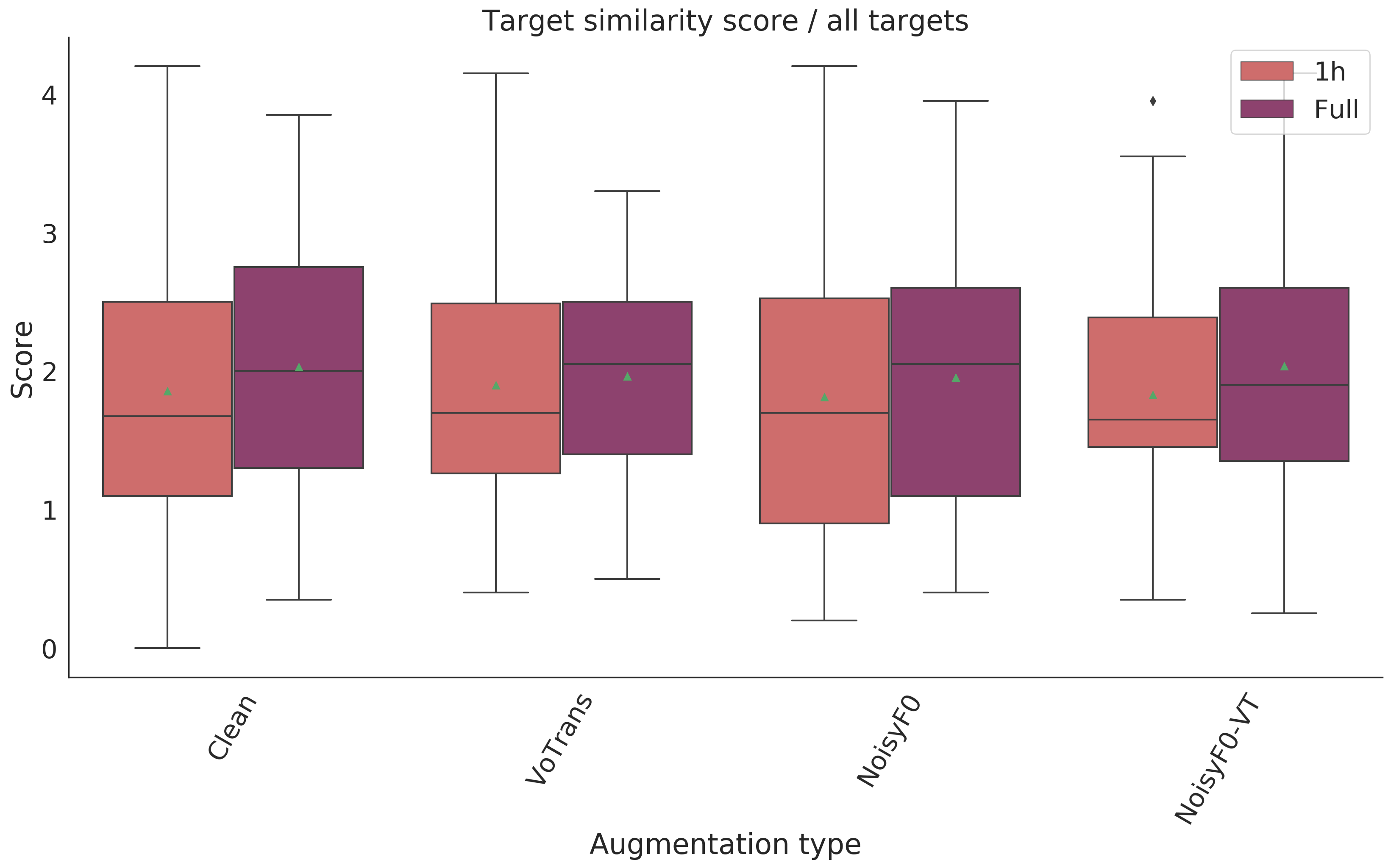}
         \caption{Similarity score subjective listening test results for all targets.}
         \label{fig:sim_all}
     \end{subfigure} \hfill
    \begin{subfigure}{0.65\columnwidth}
         \includegraphics[width=\columnwidth]{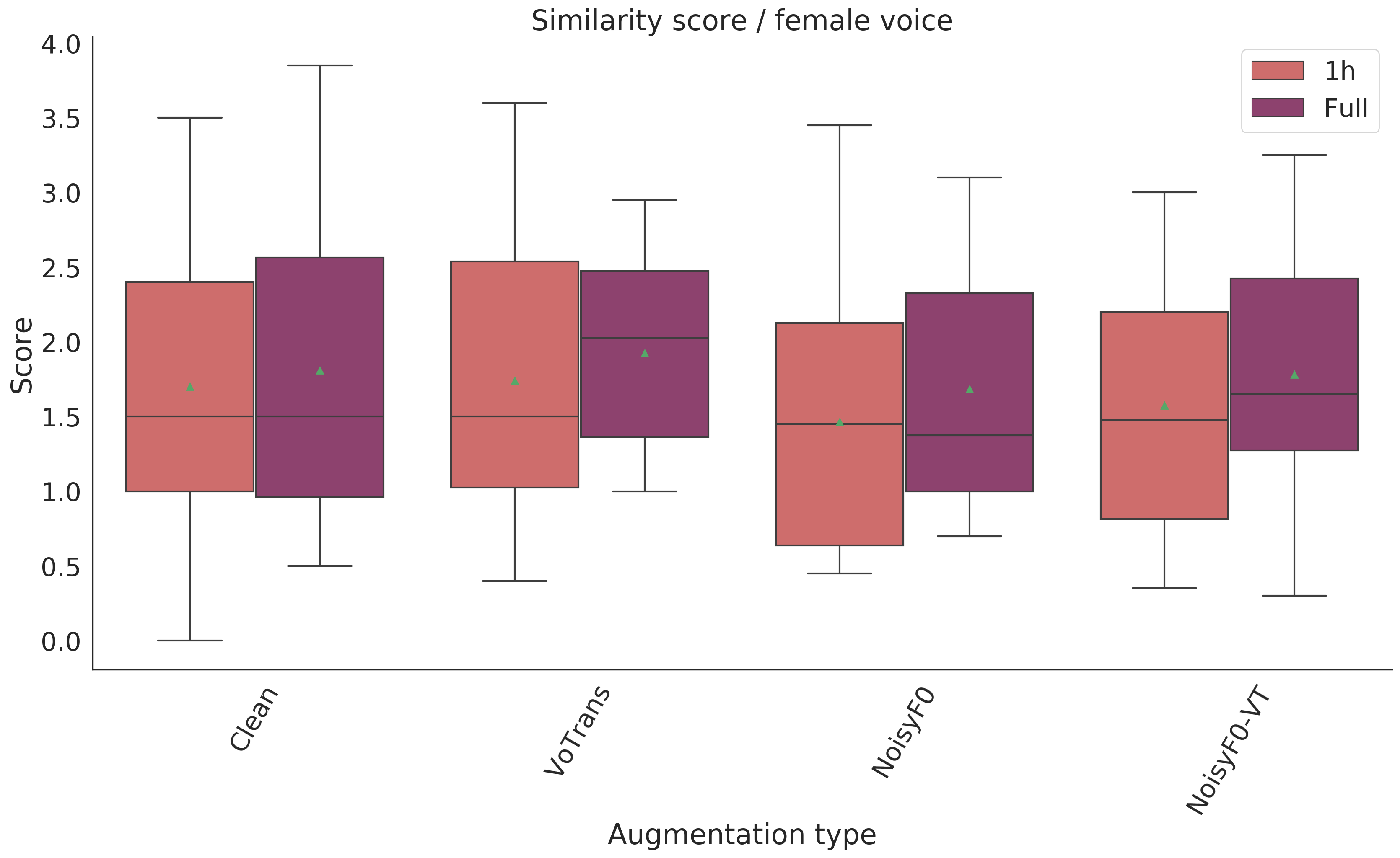}
         \caption{Similarity score subjective listening test results for the female target.}
         \label{fig:sim_female}
     \end{subfigure} \hfill
    \begin{subfigure}{0.65\columnwidth}
         \includegraphics[width=\columnwidth]{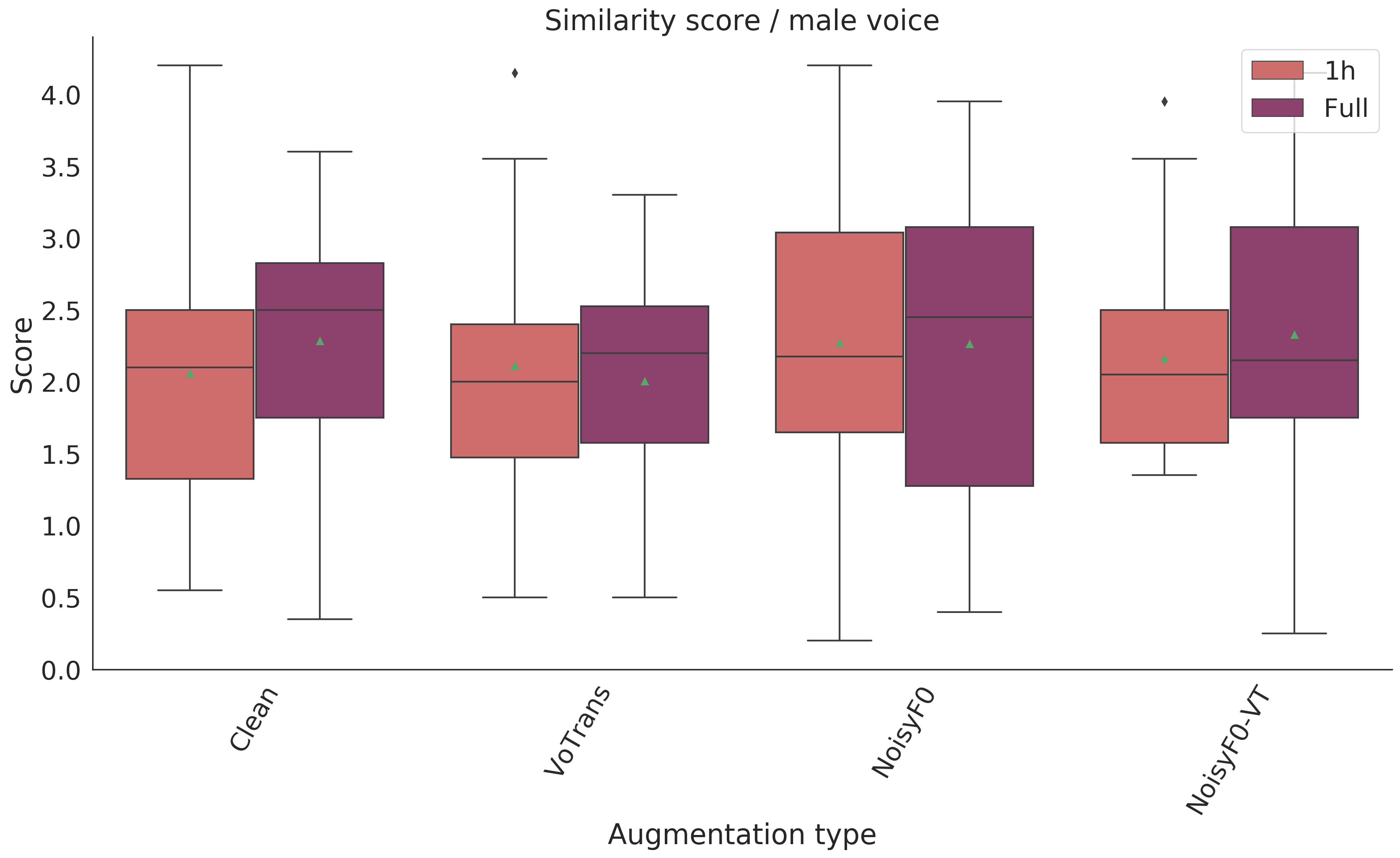}
         \caption{Similarity score subjective listening test results for the male target.}
         \label{fig:sim_male}
     \end{subfigure} \hfill
  \caption{Similarity score subjective listening test results for all data augmentations schemes and all targets. }
  \label{sim}
\end{figure*}

\end{document}